\documentclass[a4paper]{article}
\usepackage{balance} 
\usepackage{INTERSPEECH2022}
\usepackage{algorithm}
\usepackage{algpseudocode}
\usepackage{cite}

\title{Adversarial Speaker Distillation for Countermeasure Model on \\ Automatic Speaker Verification}
\name{Yen-Lun Liao*, Xuanjun Chen*, Chung-Che Wang, Jyh-Shing Roger Jang\thanks{* \textbf{Equal contributions.}}}
\address{Department of Computer Science and Information Engineering, National Taiwan University}
\email{\{r09922047, r09922165\}@ntu.edu.tw, \{geniusturtle, jang\}@mirlab.org}

\begin{document}

\maketitle
\begin{abstract}
The countermeasure (CM) model is developed to protect ASV systems from spoof attacks and prevent resulting personal information leakage in Automatic Speaker Verification (ASV) system. Based on practicality and security considerations, the CM model is usually deployed on edge devices, which have more limited computing resources and storage space than cloud-based systems, confining the model size under a limitation. 
To better trade off the CM model sizes and performance, we proposed an adversarial speaker distillation method, which is an improved version of knowledge distillation method combined with generalized end-to-end (GE2E) pre-training and adversarial fine-tuning. 
In the evaluation phase of the ASVspoof 2021 Logical Access task, our proposed adversarial speaker distillation ResNetSE (ASD-ResNetSE) model reaches 0.2695 min t-DCF and 3.54\% EER. ASD-ResNetSE only used 22.5\% of parameters and 19.4\% of multiply and accumulate operands of ResNetSE model.
\end{abstract}


\section{Introduction}
\label{sec:intro}

Automatic speaker verification (ASV) is a method for determining if a certain utterance is spoken by an individual. It is one of the essential biometric identification technologies widely used in real-world applications, including smartphones, smart speakers, digital wallets, etc. Through active research on various methods \cite{reynolds2000speaker,senior2014improving,lei2014novel,heigold2016end}, significant performance improvements have been created in accuracy and efficiency of ASV systems. The earliest proposed method \cite{reynolds2000speaker} used the Gaussian mixture model to extract acoustic features and then compute a score based on the likelihood ratio. End-to-end ASV models such as \cite{heigold2016end, zhang2022mfa} have been proposed to map utterances to verification scores directly. End-to-end models improve verification accuracy, making the ASV model compact and efficient. 

The ASVspoof challenge series \cite{yamagishi21_asvspoof,todisco2019asvspoof,kinnunen17_interspeech,wu2015asvspoof} were held to encourage researchers developing strong countermeasure (CM) models against spoofing audio signals such as synthetic, converted, and replayed. ASVspoof 2021 consists of three subtasks: logical access (LA), deepfake (DF), and physical access (PA). This paper focuses on the LA subtask, where the spoofing audio signals are generated from either text-to-speech (TTS) or voice conversion (VC) systems. Despite the power of today's TTS and VC technology, subtle differences still exist between spoofs and raw audio streams. A variety of CM models for LA \cite{9414234,kang21b_asvspoof,lavrentyeva2019stc,chen21_asvspoof,benhafid21_asvspoof,tak2022automatic,wu2022tackling,zeinali2019detecting,lai2019assert,tomilov21_asvspoof,wu2022partially} using GRCNNs, VGG, SE-Net, or LCNN have been proposed to detect spoofing audio signals. 
ASV and anti-spoofing models have also been proven vulnerable to adversarial attack \cite{kreuk2018fooling, liu2019adversarial}. A variety of defense methods have been proposed to improve the robustness of ASV and anti-spoofing model against adversarial attacks \cite{wu2021spotting,wu2021voting,wu2022improving,wu2021adversarial,wu2020defense,wu2020defense_2}. The biggest commonality between the generated spoofing audio and the adversarial attack audio is that they influence the decisions of the countermeasure model in imperceptible ways. 

Some ASV systems \cite{thomas2020edgedevice, granqvist2020improving} are running on edge devices, in order to avoid network transmission failure. The CM model is often used in conjunction with the ASV model. In edge devices, the models need to be lighter to account for limited computing power and storage space. Knowledge distillation (KD) \cite{hinton2015distilling} is one classical method used to reduce the model size, where the knowledge is transferred from a larger teacher model to a more lightweight student model. However, KD often degrades the performance while reducing the model size of the model. To maintain or prevent the model performance degradation after distillation, one straightforward motivation is to train a powerful teacher model firstly.

% To enable the CM system to sense the gap between the audio produced by TTS and VC technology and the real audio, we emphasize speaker separation to make the decision boundary between modified and bona fide audio for the same speaker clearer. We also utilize the adversarial example to further improve the ability of CM model to perceive the subtle difference between modified and bona fide audio, which inspired by training by adversarial example can easier perceive the subtle perturbation in ASV task \cite{wu2020defense}. In this paper, we mainly have the following contributions:

To enable the CM system to sense the gap between the audio produced by TTS and VC technology and the real audio, we separate the embeddings of different spoofing conditions and narrow those of the same spoofing condition. We also utilize the adversarial example to improve the ability of CM model to perceive the subtle difference between modified and bona fide audio, which inspired by training by adversarial example can easier perceive the subtle perturbation in ASV task \cite{wu2020defense}. In this paper, we mainly have the following contributions:

\begin{itemize}
    \item[1)] To our best knowledge, this is the first to explore the lightweight ASV spoofing CM model.
    % \item[1)] To our best knowledge, this is the first application of knowledge distillation in ASV spoofing CM model.

    \item[2)] We proposed an adversarial speaker distillation method, which combined with generalized end-to-end (GE2E) \cite{GE2E} pre-training and adversarial fine-tuning for the teacher model, and used KD to obtain the student model. 

    \item[3)] Experiments showed that our proposed training strategies effectively improved student model performance, while maintaining a balance between performance and resource consumption. 
\end{itemize}

\section{Methods}
\label{sec:training_strategy}
\begin{figure}[t]
\includegraphics[width=\columnwidth]{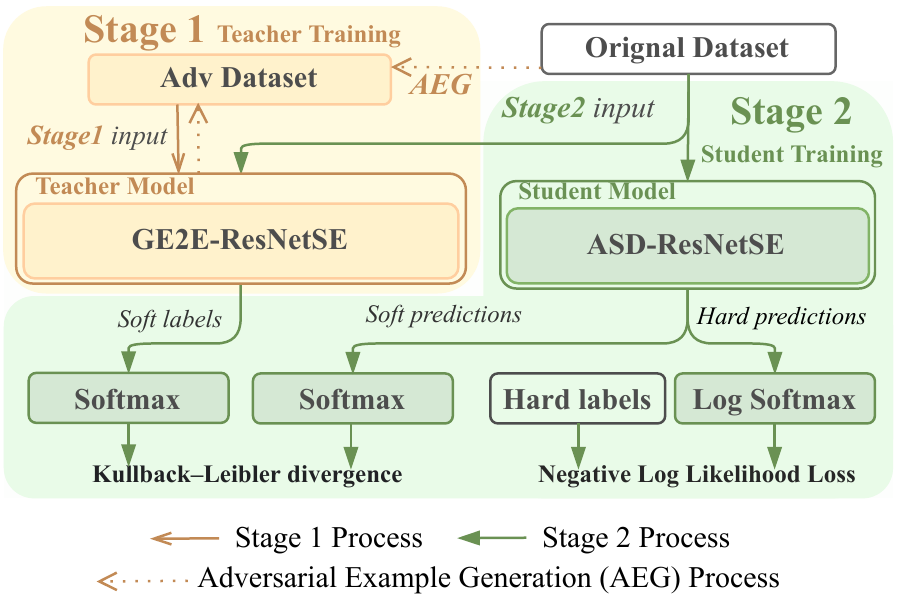}
\caption{The proposed the training strategies for lightweight CM model for ASV. ``Adv Dataset'' represents the adversarial dataset generated by the AEG process.}
\label{figure:workflow}
\end{figure}

In this paper, we designed an adversarial speaker distillation training strategy and used ResNetSE \cite{chung2020in} as the backbone model. The establishment of the teacher model involved two steps: pre-training and fine-tuning. The teacher model, GE2E-ResNetSE, is pre-trained by GE2E loss and adversarial fine-tuned by negative log-likelihood (NLL) loss.
% Moreover, adversarial example generation (AEG) was invoked before or during teacher model training. 
Note that adversarial fine-tuning means injecting an adversarial speaker class into the original dataset.
The student ResNetSE (ASD-ResNetSE) distilled adversarial speaker knowledge from GE2E-ResNetSE. Each utterance has two labels, namely the speaker and the spoofing condition. We will use them in different steps respectively. The overall process is shown in Figure \ref{figure:workflow}, and the details will be described in the following subsections.

\subsection{Model structure}
\begin{figure}[t]
\includegraphics[width=\columnwidth]{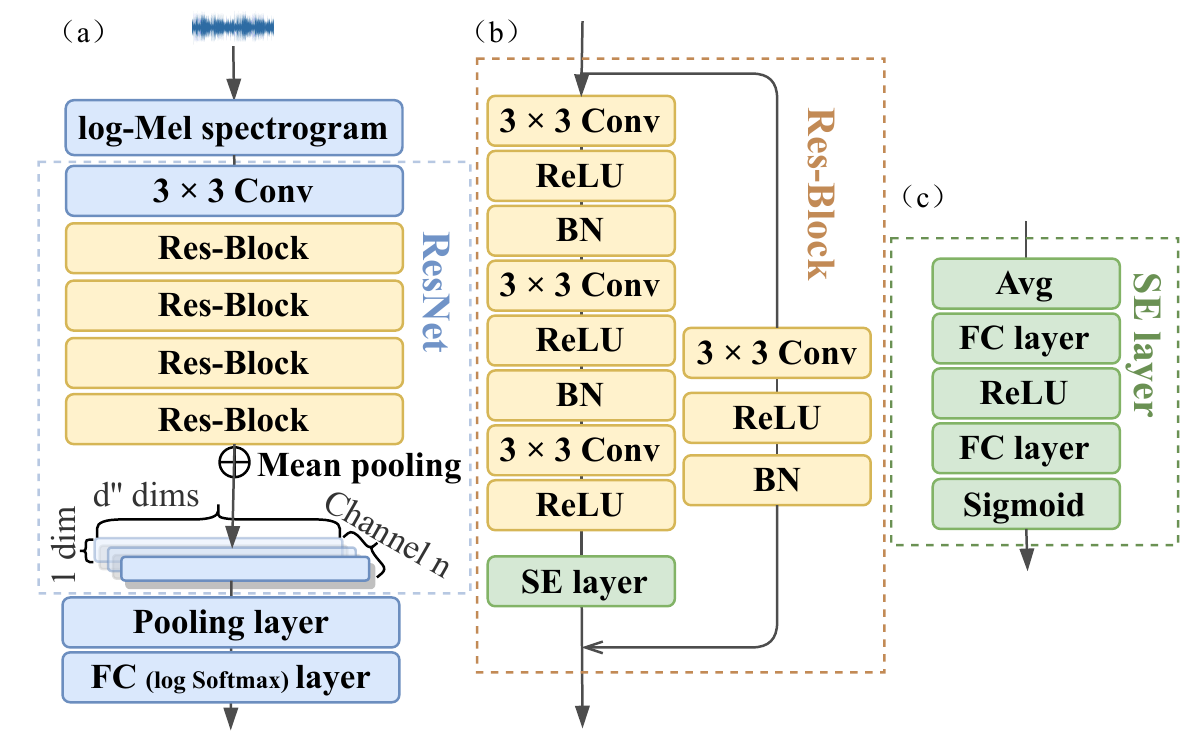}
\caption{{\it Model structure of ResNetSE. (a) Sketch of ResNetSE model. \ (b) Details of Res-Block.  \ (c) Details of squeeze-excitation layer.}}
\label{figure:ResNetSE}
\end{figure}

Excellent results have been achieved by ResNet and its improved version ResNetSE on image processing and ASV \cite{GE2E,heo2020clova,chung2020in}. Figure \ref{figure:ResNetSE} shows the structure of ResNetSE, where self-attentive pooling was used as the pooling layer to enhance model flexibility for accepting audio streams of various lengths. 
It can reduce the input signals of any length to the same dimension by using pooling in our system.
The output target of ResNetSE is an 8-element vector, indicating the input audio is bona fide, TTS spoofing methods, VC spoofing methods, and an adversarial speaker. Note that only the adversarial fine-tuning step has an adversarial speaker class. The difference between GE2E-ResNetSE and ASD-ResNetSE is the number of channels: (32, 64, 128, 256) for the former and (16, 32, 64, 128) for the latter. 

\subsection{Teacher training}

The process of training the teacher model was composed of two main steps. 
The first step of using GE2E pre-training with spoofing condition classes is to ensure the model has sufficient discriminating power between different spoofing conditions. In this step, the GE2E loss was computed using the pooling layer of ResNetSE (Figure  \ref{figure:ResNetSE} (a)).
In the second step, we fine-tuned the whole model with the ASVspoof2021-provided data and the injected adversarial data generated by AEG. This process makes the model focus on distinguishing fake audio streams from bona fide ones.
% % .Since the loss was calculated using the output layer (which was used to determine whether the input audio was bona fide), this fine-tuning step focuses the model on distinguishing fake data by using spoof labels as targets. 
% The teacher model training process was described below.

\subsubsection{Generalized End-to-End Pre-training}
\label{sec:GE2E}
GE2E loss \cite{GE2E} was proposed for speaker verification, and a variant of this approach has been used to detect replayed spoofing attacks (PA subtask of ASVspoof 2021) \cite{lei21_asvspoof}. This paper uses GE2E loss calculated according to spoofing condition classes in LA subset to obtain the initial model. Firstly, each batch included $M$ utterances from one of the $N$ different conditions. The utterances $\mathbf{x}_{nm}$ in the batch (except the query itself) formed the centroids $c_{n}$ of each condition, where $1 \leq n \leq N$, $1 \leq m \leq M$. The formula of $c_{n}$ is defined as:
\begin{equation}
c_{n} = \frac{1}{M}\sum^{M}_{m=1}\mathbf{x}_{nm}
\label{eq1}
\end{equation}

Each utterance embedding was expected to be close to its corresponding spoofing condition centroid but far from the centroids of other spoofing conditions. Thus, a similarity matrix $\mathbf{S}_{nm,k}$ was defined to describe the scaled cosine similarity of utterances with centroids $k$, where $w, b$ are learnable parameters in the expression:
\begin{equation}
\mathbf{S}_{nm,k} = w \cdot \mathrm{cos}(\mathbf{x}_{nm},c_{k}) + b
\label{eq2_v2}
\end{equation}

Softmax was applied to the similarity matrix for every category (from 1 to $N$) when calculating GE2E loss. The overall loss function was defined in equation \ref{eq3_v2}, which means that utterance embeddings of the same spoofing condition should be close to each other, and far from those of other spoofing conditions.
\begin{equation}
\mathcal{L}_{GE2E} = -\mathbf{S}_{nm,n} + \mathbf{log}\sum^{N}_{k=1}exp(\mathbf{S}_{nm,k})
\label{eq3_v2}
\end{equation}

\vspace{-0.5em}
\subsubsection{Adversarial fine-tuning}

After the teacher model was initialed by GE2E pre-training, it will be fine-tuned using NLL loss with spoofing condition label. The spoofing condition label is composed of 8 classes: 1 bona fide, 6 spoof methods, and 1 adversarial speaker. The additional adversarial speaker class is generated by the Adversarial Example Generation (AEG) algorithm described below.

% After the teacher model was pre-trained by GE2E using the output of the pooling layer, the final output indicating the 8 spoofing condition classes (1 bona fide, 6 spoof methods, 1 adversaria) was used as the target for fine-tuning. The dataset used in this stage is the original dataset plus the adversarial samples generated by static or active AEG. The loss function is NLL loss.

% \subsubsection{Adversarial Example Generation}

AEG was a process that deliberately generated a tiny perturbation to the original sample to generate adversarial audio. This work adopted the basic iterative method (BIM) \cite{kurakin2016adversarial} for AEG. The audio input of the AEG algorithm was a same-speaker randomly-selected bona fide audio signals $\mathsf{W_1}$ and $\mathsf{W_2}$, and the output was the newly generated sample. Extract\_Feature($\mathbf{M}$, $\mathsf{W}$) meant using model $\mathbf{M}$ to extract features for $\mathsf{W}$. The score $s$ indicates the similarity between the forged and the original audio. Only the new sample with $s$ smaller than $threshold$ will be used as an adversarial sample. The algorithm is detailed in Algorithm \ref{algrithm:BIM}. Based on this, we implemented two AEG methods, static AEG and active AEG.

\noindent\textbf{Static AEG.} Static AEG used the GE2E pre-trained model as the input for Algorithm \ref{algrithm:BIM} to generate attack data. All the generated attack data will be relabeled as an adversarial example and injected into the original data set. Since static AEG created all adversarial samples before fine-tuning, it did not increase the overhead of fine-tuning. 

\noindent\textbf{Active AEG.} The difference between static AEG and active AEG was that active AEG executed BIM before each epoch, which means that the same input audios will produce different adversarial samples in different epochs. Although this additional work may increase training time, active AEG is expected to generate more architecture-specific data.

\label{sec:AEG}
\begin{algorithm}
\caption{Basic Iterative Method (BIM)}
\label{algrithm:BIM}
\begin{algorithmic}[1]
\Require{Two audios, $\mathsf{W_1}$ and $\mathsf{W_2}$ belonging to the same speaker. Model $\mathbf{M}$.} $\alpha$, $iter$, and $threshold$ are controllable variables.
\Ensure{Feature of the forged audio.} 
\Statex
\State $\mathbf{X_1} \gets$ Extract\_feature($\mathbf{M}, \mathsf{W_1}$)
% \State $\mathbf{D_{max}} \gets$  zero\_array( len($\mathsf{W_1}$) )
\State $\mathbf{D} \gets$  Zero\_array(len($\mathsf{W_1}$))
\For{$t \gets 1$ to $iter$}
    \State $\mathbf{X_2} \gets$ Extract\_feature($\mathbf{M}, \mathsf{W_2} + \mathbf{D}$)
    \State s $\gets$ Cos-similarity($\mathbf{X_1}, \mathbf{X_2}$)
    % \State $d \gets \nabla_{\mathbf{D}}(s)$
    \State $\mathbf{D} \gets$ clip $(\mathbf{D}$ + $\alpha$ * sign($\nabla_{\mathbf{D}}(s)))$
\EndFor
\State $\mathbf{X_2} \gets$ Extract\_feature($\mathbf{M}, \mathsf{W_2} + \mathbf{D}$)
\State s $\gets$ Cos-similarity($\mathbf{X_1}, \mathbf{X_2}$)
\If{s $\leq threshold$ }
    \State \Return NULL
\EndIf
\State \Return $\mathbf{X_2}$
\end{algorithmic}
\end{algorithm}

% \vspace{-0.5em}
% \subsubsection{Adversarial fine-tuning}

% In this step, teacher model was initialed by GE2E pre-training, as shown in section \ref{sec:GE2E}.

% We utilize AEG to generate an additional class named adversarial speaker.  

% % After the teacher model was pre-trained by GE2E using the output of the pooling layer, the final output indicating the 8 spoofing condition classes (1 bona fide, 6 spoof methods, 1 adversaria) was used as the target for fine-tuning. The dataset used in this stage is the original dataset plus the adversarial samples generated by static or active AEG. The loss function is NLL loss.

\subsection{Student training}
This stage used KD loss \cite{hinton2015distilling} to transform the capabilities of GE2E-ResNetSE into ASD-ResNetSE. KD loss mainly consists of Kullback–Leibler divergence (KL) and NLL loss. KL loss is to lets the student learn soft targets from the teacher's output.
 NLL loss enables the student to learn the hard spoof condition label. The overall loss function is:
\begin{equation}
\mathcal{L}_{KD} = \gamma T^{2} \times KL (\frac{O_{s}}{T}, \frac{O_{t}}{T}) + (1-\gamma)\mathcal{L}_{NLL}
\label{eq4}
\end{equation}
where $O_{s}$ means the output of the student model, $O_{t}$ means the output of the teacher model, $\mathcal{L}_{NLL}$ is the NLL loss between the prediction of students and ground truth classes, and $T$ is the parameter controlling the distillation temperature, and $\gamma$ is the weight for balancing the contribution from the teacher and the ground truth class. After completing this stage, the ASD-ResNetSE will be used to measure the final model performance.

\section{Experiment Setup}
\label{sec:exp}
\textbf{Dataset and evaluation metrics.} All experiments follow the ASVspoof 2021 \cite{yamagishi21_asvspoof} settings, and were performed on the ASVspoof 2019 \cite{todisco2019asvspoof} (for validation) and ASVspoof 2021 (for evaluation) LA dataset. We only reported the 2021 results here. Training and development partitions of ASVspoof 2019 were used to construct the countermeasures' system. The system was evaluated on the ASVspoof 2019 and ASVspoof 2021 evaluation sets. There were 7 spoofing categories, including bona fide, A01-04 spoofed by TTS, and A05-06 spoofed by VC. Meanwhile, all the audio files were labeled with speakers. The training set includes 20 speakers, while the development set includes 10 speakers. The dataset includes a training set of 25,380 utterances, a development set of 24,856 utterances, and an evaluation set of 181,566 utterances. Minimum tandem decision cost function (min t-DCF) \cite{kinnunen2020tandem} and equal error rate (EER) were used to evaluate the effectiveness of the countermeasure (CM) models. 

\noindent\textbf{Waveform augmentation.} The number of utterances in the evaluation data set was much larger than in training or developing sets. Therefore, we used waveform augmentation to expand the training data to increase system robustness. First, we randomly selected music, voice, or noise in MUSAN \cite{snyder2015musan} and trimmed or padded it to the same length as the target utterance, then added it to the audio file to generate new audio. Next, we randomly selected audio from the RIR noise data \cite{rir-filter} set and convoluted it with the target audio to generate new audio for reverberation simulation of different room sizes. The training data of all models in the experiment were augmented using the above-mentioned on-the-fly methods.

\noindent\textbf{Training Details.} 
We extracted 40 dimensions log-Mel spectrogram with a 25 ms window size, a 10 ms hop size, and an FFT size set as 512 as the input, while all audio files had a sample rate of 22,050 Hz. Following acoustic extraction, we applied instance normalization to the feature. We set $\alpha = 3.0, iter = 5, threshold = 0.4$ in AEG augmentation. The $\alpha$ and $iter$ are referred from the empirical value of \cite{wu2020defense}. We applied waveform augmentation and used the Adam optimizer during the end-to-end teacher and student model training. In the beginning, the learning rate is 0.0003, and every two epochs will become 0.95 times the original. In KD loss, $\gamma = 0.5, T = 5$.

\section{Results and Analysis}
\label{sec:results}
\subsection{Ablation Study}
% In this section, we will perform ablation study of GE2E, pre-training, AEG, Knowledge distillation.

\textbf{GE2E.} We compared the performance of ResNetSE and GE2E-ResNetSE model in Table \ref{table:ge2e}. The results show that GE2E pre-training reduces min t-DCF score from 0.3143 to 0.3003 and reduces EER from 5.78 to 5.10. GE2E pre-training enables the model to distinguish the information of the same spoofing condition from the information of other spoofing conditions, which helps classify the spoofing classes by providing additional spoofing condition information. Rather than classifying all the data at once, GE2E enabled the classification of spoof and non-spoof starting from classifying a particular spoofing condition case, which was expected to have better results. 

% So G-ResNetSE will use GE2E pre-training by default in the following experiments.

\begin{table}[ht]
\renewcommand\arraystretch{1.2}
\centerline{
\begin{tabular}{ccccc}
\hline
\hline
\textbf{Model} & \textbf{Loss} & \textbf{min t-DCF} & \textbf{EER} \\
\hline
\hline
ResNetSE & $\mathcal{L}_{NLL} $ &  0.3143 & 5.78 \\
GE2E-ResNetSE & $\mathcal{L}_{GE2E}$ + $\mathcal{L}_{NLL} $ &  0.3003 & 5.10 \\
\hline
\end{tabular}}
\vspace{1mm}
\caption{{\it Influence of GE2E pre-training.}}
\label{table:ge2e}
\vspace{-5mm}
\end{table}

\begin{table}[ht]
\renewcommand\arraystretch{1.2}
\centerline{
\begin{tabular}{lccc}
\hline
\hline

\textbf{Teacher Model} & \textbf{AEG} & \textbf{min t-DCF} & \textbf{EER}  \\
\hline
\hline
(A) ResNetSE & None   & 0.3143 & 5.78 \\
(B) GE2E-ResNetSE & None   & 0.3003 & 5.10  \\
\hline
(C) GE2E-ResNetSE & Static & 0.2931 & 5.05  \\
(D) GE2E-ResNetSE & Active & \textbf{0.2869} & \textbf{4.59}\\
\hline
\hline
\textbf{Student Model} & \textbf{AEG} & \textbf{min t-DCF} & \textbf{EER}  \\
\hline
\hline
(A$'$) ASD-ResNetSE & None   & 0.2987 (4.9\%) & 4.83 (16.4\%) \\
(B$'$) ASD-ResNetSE & None   & 0.2826 (5.8\%) & 4.44 (12.9\%) \\
\hline
(C$'$) ASD-ResNetSE & None   & \textbf{0.2695} (8.0\%) & \textbf{3.54} (29.9\%) \\
(D$'$) ASD-ResNetSE & None   & 0.2903 (-1.1\%) & 4.76 (-3.7\%) \\
\hline
\end{tabular}}
\vspace{1mm}
\caption{{\it Comparison of adversarial example generation methods and knowledge distillation. (A$'$)-(D$'$) denote student model obtained by distillation of corresponding teacher (A)-(D). ($\cdot$\%) refers to the relative  improvement ratio between student and teacher.}}
\label{table:adversarial}
\vspace{-8mm}
\end{table}

\noindent\textbf{AEG.} 
The results (B)(C)(D) in the upper part of Table \ref{table:adversarial} show that both active and static AEG improved the performance of GE2E-ResNetSE. In particular, the active method effectively reduced the min t-DCF from 0.3003 to 0.2869 and the EER from 5.10 to 4.59\% through on-the-fly augmentation. It showed that adding model-weakness data to the dataset during training can help the model perceive the subtle differences inside the spoofing audio streams. From the perspective of data preparation, the active method is easier to implement, whereas the static method generates the whole dataset before training.

\noindent\textbf{Knowledge Distillation.} The lower half of Table \ref{table:adversarial} is the result of ASD-ResNetSE obtained by distillation of the corresponding teacher through the KD process. The result of (A$'$) is our knowledge distillation baseline. Its teacher (A) did not use GE2E pre-training and adversarial fine-tuning. The teacher (B) only used GE2E pre-training. The teachers (C) and (D) of (C$'$) and (D$'$) are based on GE2E-ResNetSE and also use static AEG and active AEG to inject adversarial speaker class during fine-tuning, respectively. Note that we don't use AEG during the distillation process.

In Table \ref{table:adversarial}, (A$'$) gains 4.9\% min t-DCF and 16.4\% EER improvement after distillation. (C$'$) obtains 8.0\% min t-DCF and 29.9\% EER improvement combined with GE2E pre-training and static AEG fine-tuning, which outperforms baseline (A$'$). However, (D$'$) degrades -1.1\% min t-DCF and -3.7\% EER. In most distillation results, the performance of the student model did not decrease but increased. Previous work \cite{frankle2018lottery} has found that subnetworks exist in the original network that can reach or exceed the original model performance. Wang \cite{wang2022lighthubert} has tried to find such a subnetwork through KD. The above result shows that static AEG is more helpful than active AEG in finding a super subnetwork in the adversarial speaker distillation process.

\subsection{Overall Performance}

\begin{figure}
\centering
\includegraphics[width=5.5cm]{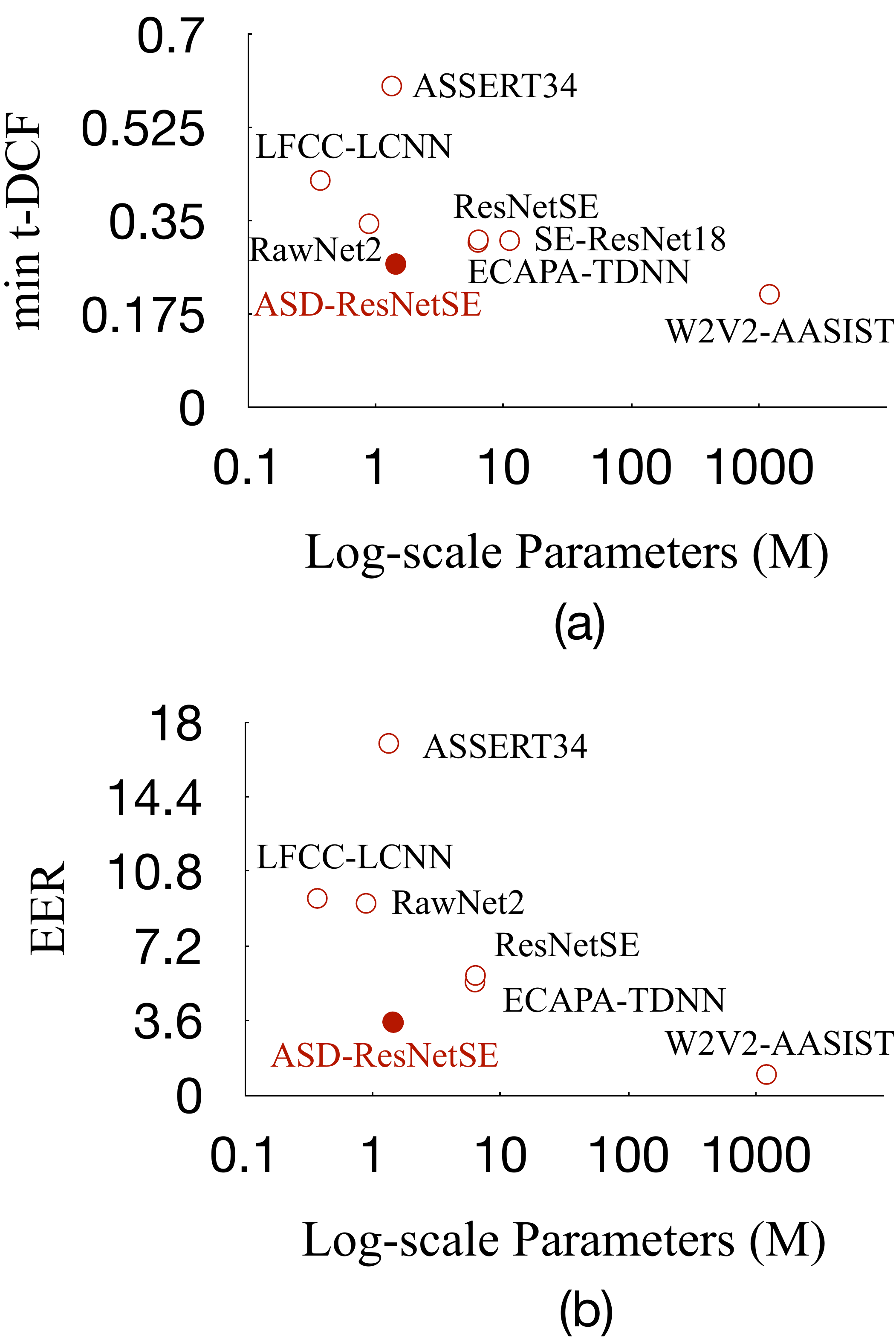}
\caption{Overall performance comparison.}
\label{fig:OverallPer}
\vspace{-1.5em}
\end{figure}

To further analyze overall performance, we select our best ASD-ResNetSE model (C$'$) and ResNetSE (A) and the other existing countermeasure models, such as RawNet2 \cite{9414234}, SE-ResNet18 \cite{kang21b_asvspoof}, LFCC-LCNN \cite{lavrentyeva2019stc}, ECAPA-TDNN \cite{chen21_asvspoof}, ASSERT34 \cite{benhafid21_asvspoof} and W2V2-AASIST \cite{tak2022automatic}. The best ASD-ResNetSE model and GE2E-ResNetSE model were from Table \ref{table:adversarial}. For a fair comparison, we only consider single models instead of fusion models. Figure \ref{fig:OverallPer} visualize the relation between the model sizes and their performance of min t-DCF, EER. Because SE-ResNet18 does not provide EER results, so we do not make a comparison here.

We mainly compare three aspects in Figure \ref{fig:OverallPer}.
Firstly, ASD-ResNetSE not only outperforms most countermeasure models but also has only a 22.5\% model size of ResNetSE. 
Secondly, ASD-ResNetSE is more than 50\% min t-DCF lower than ASSERT34, which has a similar model size to ASD-ResNetSE. Moreover, the EER of ASD-ResNetSE is only about 20\% of the EER of ASSERT34. The above results demonstrate that ASD-ResNetSE can  obtain a better trade-off between the model size and performance. 
Thirdly, W2V2-AASIST achieves state-of-the-art min t-DCF and EER, consisting of a self-supervised learning frontend wav2vec2.0 \cite{baevski2020wav2vec, xu2021self} and a countermeasure backend AASIST \cite{jung2022aasist}. However, its parameters are over 1200 megabytes, so this is not practical for using it on edge devices.

\subsection{Model Size and Operands}

We compared the size and multiply and accumulate operands (MACs) of several methods, shown in Table \ref{table:statistic}. ASD-ResNetSE, with only 0.90G MACs, had a significantly smaller model size than ResNetSE. As the model size, ASD-ResNetSE also had similar MACs with ASSERT34. Combining the results in Table \ref{table:statistic} and Figure \ref{fig:OverallPer}, ASD-ResNetSE stands out for its capacity, efficiency and effectiveness.
%was fast, small and powerful. 

\begin{table}[ht]
\renewcommand\arraystretch{1.2}
\centerline{
\begin{tabular}{ccccc}
\hline
\hline
Model & Param & MACs & min t-DCF & EER\\
\hline
\hline
ResNetSE & 6.39 M & 4.62 G &  0.3143 & 5.78 \\
ASSERT34 \cite{benhafid21_asvspoof} & \textbf{1.34 M} & 1.01 G & 0.603 & 16.98 \\ 

ASD-ResNetSE & 1.44 M & \textbf{0.90 G} & \textbf{0.2695} & \textbf{3.54}\\
\hline
\end{tabular}}
\vspace{1mm}
\caption{\label{table:statistic} {\it Comparison of model size and operands.
Param represents model parameters. MACs represents multiply and accumulate operands.}.}
\vspace{-8mm}
\end{table}

\section{Conclusion}
\label{sec:conclusion}

This paper is the first to explore the lightweight CM model for ASV and propose an adversarial speaker distillation method, an improved version of knowledge distillation. In the evaluation phase of the ASVspoof 2021 Logical Access task, our ASD-ResNetSE reaches 0.2695 min t-DCF and 3.54\% EER with only used 22.5\% of parameters and 19.4\% of MACs of the original ResNetSE model. 
The experiment results demonstrate that ASD-ResNetSE stands out for its capacity, efficiency and effectiveness.

\section{Acknowledgements}

% The task was completed thanks to the instructions of E-sun and the hardware support of the Taiwan computational cloud (TWCC).
Thanks for technical support from Chi-Han Lin, E.SUN COMMERCIAL BANK, LTD. This work was partially supported by the Ministry of Science and Technology, Taiwan (Grant no. MOST109-2221-E-002-163-MY3).
% \newpage
\balance

\bibliographystyle{IEEEtran}
\bibliography{mybib}

\end{document}